# New Models for $UO_2$ Fuel Structure Evolution under Irradiation in Fast Reactors

*M.S. Veshchunov**

Nuclear Safety Institute (IBRAE), Russian Academy of Sciences

52, B. Tulskaya, Moscow, 115191, Russia

Moscow Institute of Physics and Technology (MIPT) (State University)

9, Institutskii per., Dolgoprudny, Moscow Region, 141700, Russia

**Abstract**

On the base of analysis of experimental observations and critical assessment of existing models for oxide fuel structure evolution under operation conditions of fast reactors, new models for fuel restructuring and coring are proposed. The restructuring model describes coherent motion in the temperature gradient of various voids (gas bubbles, sintering pores and large lenticular pores) and grain boundaries, to which the voids are attached. As a result, the model explains elongation of thermally growing equiaxed grains and formation of columnar grains, and predicts a rapid formation of extended columnar grain zone during a relatively short initial period of fast reactor irradiation. The coring model describes formation and growth of the central void in the fuel pellet, activated by mass transport from the inner to the outer zone of the pellet under stresses induced by inhomogeneous fuel densification in the initial period of irradiation.

***** Corresponding author:
Phone: +7 (495) 955 2218, fax: +7 (495) 958 0040
E-mail: vms@ibrae.ac.ru

# 1 Introduction

Fast reactor oxide fuel is generally subjected to steady-state and transient power operating conditions leading to significant irreversible changes in the fuel structure. When cylindrical oxide fuel pins are placed in a neutron flux, the volumetric heating rates and low thermal conductivity of the fuel combine to produce high temperatures and very steep radial thermal gradients. These conditions can lead to the phenomenon commonly referred to as restructuring. The most distinct macroscopic aspects are divided into columnar, equiaxed and unrestructured (as-fabricated) fuel zones. The basic physical processes that produce these zones have been identified as grain growth kinetics for the equiaxed zone, and porosity migration for the columnar zone.

In fast reactors this restructuring affects the fuel performance in many ways (fission gas release, swelling, actinide, oxygen and fission product redistribution) and involves important matter transport processes within the fuel. The mechanism of this restructuring has been studied for many years and it was concluded that formation of columnar grains is related to the migration of pores with a particular shape, elongated in the direction perpendicular to the thermal gradient [1-3]. These pores, which have been called 'lenticular', are thought to be formed in fuel elements from cracks [3, 4] and their migration is thought to be mainly responsible for the formation of columnar grains in the central hot fuel region. Solid state diffusion mechanisms (grain boundary migration) were usually assumed to play a less important role in the generation of the columnar grain structure, despite interaction of pores with a grain boundary generally induces strong retarding effect for their motion. In particular, in the traditional models it is (implicitly) assumed that relocation velocity of pores is not influenced by grain boundaries, to which these pores are attached, this requires additional justification. Self-consistent consideration of coherent motion in the temperature gradient of lenticular pores and grain boundaries to which pores are attached is thus an important problem that allows calculating elongation of grains and evaluating the size of the columnar zone. This problem will be considered in Section 2.

Lenticular pores have also been thought to collect at the centre line of the fuel, giving rise to the formation and growth of a central void [5]. Such central voids have been observed in irradiated fuel elements and the phenomenon has been termed "coring". The large radial thermal gradient existing in reactor fuel pins at power produces a gradient across a pore. The atoms on the hotter surface evaporate, move across the pore, and condense on the cooler surface. This causes the pore to move with the velocity proportional to the thermal gradient.

However, in a cylindrical fuel rod with an axisymmetric power distribution, the thermal gradient is practically zero at the centre of the pin, or inner surface if a central void exists (owing to zero heat flux condition at the centre). Therefore, the pore velocity decreases along with the temperature gradient in the inner zone and tends to zero at the void surface, significantly retarding pores sinking into the void and resulting in accumulation of pores in the innermost cell. To overcome this situation, in the existing models the thermal gradient at the inner fuel cell boundary is artificially assumed to be the average value across the central cell. In order to avoid such a nonphysical treatment, an additional mechanism of the central void growth will be proposed in Section 3. This mechanism associated with rapid fuel densification in the hot central zone (owing to thermal shrinkage of as-fabricated porosity), can be considered as complementary to the traditional mechanism and applied in a conservative approach for a reasonable interpretation of available observations.

The main outcomes are formulated in Section 4.

# 2 Model for grain elongation and columnar structure formation

## 2.1 Pore migration mechanisms

There are three mechanisms that could lead to the pore migration in the fuel: (1) evaporation-condensation across the pore, (2) pore surface diffusion, and (3) mass diffusion around the pores. The mass diffusion process is assumed to be negligible because of the high activation energy required to make cations in a solid sufficiently mobile to produce appreciable mass transport. If the pores are small and at lower temperatures, the surface diffusion process would be expected to dominate. For large pores and at high temperatures the evaporation-condensation process is expected to be the dominant process.

For the surface diffusion mechanism the velocity of a pore of radius $R$ in the temperature gradient $(\nabla T)_{matr}$ in a fuel matrix is calculated as [6]

$$v_{sd} = \frac{2D_s\delta}{R}\frac{\alpha_s}{T}(\nabla T)_{matr} \equiv \beta_p^{(s)}(\nabla T)_{matr}, \tag{1}$$

where $D_s \approx 50\cdot\exp(-54126/T)$ m$^2$/s is the surface self-diffusivity of uranium atoms [7], $\alpha_s$ is the thermodiffusion ratio, $\delta \approx 0.5$ nm is the surface layer thickness.

For the evaporation mechanism, when mass transport is controlled by the vapour diffusion in the pore, the pore velocity is calculated as [8, 9]

$$v_p = (1+\chi)\frac{\Omega D_{fg}}{k}\left(\frac{\partial(P_f/T)}{\partial T} + \frac{\alpha_f}{T}P_f\right)(\nabla T)_{matr}, \tag{2}$$

where $P_f$ is the $UO_2$ vapour partial pressure in the pore, $D_{fg} = \frac{3}{8\pi(n_f+n_g)d_m^2}\sqrt{\frac{\pi kT}{2m^*}}$ is the vapour diffusivity in the gas mixture, $d_m$ is arithmetic mean of the diameters of the fuel vapour and gas molecules ($\approx 4.33\cdot10^{-10}$ m), $m^*$ is equivalent mass ($1/m^* \approx (1/m_g + 1/m_f)\cdot 6.023\cdot10^{26}$ kg$^{-1}$), $n_f$ is the number density of fuel vapour molecules filling a pore, $n_g$ is the number density of gas molecules filling a pore, $\Omega$ is the atomic volume in the fuel matrix. The form factor $\chi$ is equal to 0.5 for spherical pores and $(\chi+1)\approx l/d$ for lenticular pores with $l >> d$ [10], where $l$ and $d$ are the longitudinal and transverse sizes of a pore. Since the thermodiffusion ratio $\alpha_f$ is proportional to the vapour concentration in the gas mixture, $\alpha_f = \beta c(1-c)$, where $c = n_f/(n_f+n_g)$, $\beta \approx (m_f - m_g)/(m_f + m_g)$, the corresponding term in Eq. (2) can be neglected owing to an extremely small value of the vapour partial pressure,

$$v_p \approx (1+\chi)\frac{\Omega D_{fg}}{k}\frac{\partial(P_f/T)}{\partial T}(\nabla T)_{matr} \equiv \beta_p^{(vap)}(\nabla T)_{matr}. \tag{2a}$$

However, being captured by a grain boundary, a pore moves generally much slower owing to the drag force exerted by this boundary on the pore [11, 12]. For large lenticular and as-fabricated sintering pores this retarding effect is usually ignored, that requires a thorough consideration and additional justification.

## 2.2 Interactions of pores with grain boundaries

Moving in a fuel grain upward the temperature gradient, a pore reaches a grain boundary and interacts with it, since the attachment of a pore reduces the total boundary energy. Therefore, a significant force $F_m$ will be necessary to detach the pore from the boundary. If the pore has a high mobility in the temperature gradient, it can entrain the attached boundary, until the drag force $F$ exceeds the critical value $F_m$.

The critical force exerted by a spherical pore of radius $R$ on the boundary can be evaluated as [11, 9]

$$F_m \approx \pi \gamma R, \tag{4}$$

where $\gamma$ is the grain boundary tension, that in $UO_2$ is estimated as $\approx 1$ J/m$^2$. A similar relationship with a somewhat different numerical factor can be apparently used for lenticular pores.

In the case $F < F_m$, when the pore is attached to and relocates coherently with the boundary (as a pore-boundary complex), its velocity diminishes and can be calculated by consideration of superposition of diffusion fluxes in the temperature gradient and in the stress field induced by the force $F$, exerted on the pore

$$v_p = \beta_p (\nabla T)_{matr} - b_p F, \tag{5}$$

where $b_p$ is the pore mobility under external force stresses, that similarly to $\beta_p$ is controlled by one of the three mechanisms: pore surface diffusion ($b_p^{(s)}$), or mass diffusion around the pore ($b_p^{(vol)}$), or evaporation-condensation across the pore ($b_p^{(vap)}$) [6, 12, 9]:

$$b_p^{(s)} = \frac{D_s \delta}{kT} \frac{3\Omega}{2\pi R^4}, \tag{6a}$$

$$b_p^{(vol)} = \frac{D_{vol}}{kT} \frac{3\Omega}{2\pi R^3}, \tag{6b}$$

$$b_p^{(vap)} = \frac{\Gamma}{4\pi} \frac{D_{fg}\Omega}{R^3} \left( \frac{P_f \Omega}{kT} \right), \tag{6c}$$

where $\Gamma = \frac{1}{3} \cdot \frac{1+\nu}{1-\nu}$ and $\nu$ is the Poisson ratio.

On the other hand, the velocity of the grain boundary entrained by the pore with the force $F$ (of the same absolute value but acting in the opposite direction) is equal to

$$v_{gb} = b_{gb} F, \tag{7}$$

where

$$b_{gb} = \frac{D_{gb}\Omega}{wkTS} \tag{8}$$

is the mobility of a grain boundary with the surface area $S$ [13], $w \approx 1$ nm is the boundary thickness, $D_{gb}$ is the self-diffusion coefficient across the grain boundary.

In the case of $n_p$ similar pores attached to the boundary, the boundary velocity is

$$v_{gb} = n_p b_{gb} F. \tag{9}$$

The boundary moves coherently with the pores, if

$$v = v_{gb} = v_p \, . \tag{10}$$

The value of $D_{gb}$ in $UO_2$ was evaluated in [11] from the grain growth kinetics at relatively low temperatures $T \le 1700$°C as $D_{gb} \approx 4 \cdot 10^{-6} \cdot \exp(-44200/T)$ m$^2$/s, that is ≈ 3 orders of magnitude higher that the bulk uranium self-diffusivity, $D_{vol} \approx 4 \cdot 10^{-11} \cdot \exp(-35250/T)$ m$^2$/s, measured in the similar temperature range of 1400–1700°C [14]. However, being interpolated to higher temperatures of interest (say $T \approx 2000$°C), this relationship gives $D_{gb} \approx 1.5 \cdot 10^{-14}$ m$^2$/s, that is only one order of magnitude higher than the bulk self-diffusivity, $D_{vol} \approx 5.82 \cdot 10^{-9} \cdot \exp(-36638.56/T)$ m$^2$/s, directly measured in the temperature range 1900-2150°C [15]. Therefore, at high temperatures the extrapolated value of $D_{gb}$ seems to be underestimated and will be further considered as the lower limit in this temperature range.

Besides, this value of $D_{gb}$ at $T \approx 2000$°C is four orders of magnitude smaller than the longitudinal self-diffusivity of U atoms (along the grain boundary), $D_{gb}^{(l)} \approx 5.2 \cdot 10^{-6} \cdot \exp(-23769/T)$ m$^2$/s, also measured in [15] at 1900-2150°C, whereas comparison of $D_{gb}$ with the longitudinal diffusivity $D_{gb}^{(l)} \approx 4 \cdot 10^{-6} \cdot \exp(-35250/T)$ m$^2$/s, measured at temperatures ≤ 1700°C [14] (i.e. in the same temperature interval as $D_{gb}$), shows that their ratio normally attains ≈ 2 orders of magnitude. Extrapolation of this ratio to the high temperature interval $T \approx$ 1900-2150°C gives the correlation $D_{gb} \approx 5.2 \cdot 10^{-8} \cdot \exp(-23769/T)$ m$^2$/s, that exceeds the measured bulk value $D_{vol}$ by ≈ 3 orders of magnitude, and thus will be further used as the upper limit for $D_{gb}$ in this temperature range.

Solution of Eqs. (5) - (10) determines the drag force

$$F = \frac{\beta_p (\nabla T)_{matr}}{n_p b_{gb} + b_p} \, , \tag{11}$$

and relocation velocity of the grain boundary with attached pores

$$v = \frac{n_p b_{gb}}{n_p b_{gb} + b_p} \beta_p (\nabla T)_{matr} \, . \tag{12}$$

The obtained solutions, Eqs. (11) and (12), are valid only under the additional condition $F < F_m$, that after substitution of Eqs. (4) and Eq. (11) takes the form

$$F = \frac{\beta_p (\nabla T)_{matr}}{n_p b_{gb} + b_p} < \pi \gamma R \, . \tag{13}$$

For small pores with $b_p / n_p >> b_{gb}$ (corresponding to $R << n_p^{-1} \cdot 10^{-6}$ m for boundaries with radius $R_{gb} \approx 10$ μm and with the upper limit value for $D_{gb}$) Eq. (13) is reduced to $F \approx \frac{\beta_p (\nabla T)_{matr}}{b_p} < \pi \gamma R$. Since for small pores both mobilities $\beta_p$ and $b_p$ are controlled by the pore surface mechanism, Eqs. (1) and (6a), at $T \approx 2273$ K and the typical temperature gradient $\approx 2 \cdot 10^5$ K/m the obtained restriction on the pore size, $R \le 10^{-6}$ m, turns to be less strict than the original condition $b_p / n_p >> b_{gb}$. However, the retarding effect is significant and the boundary entrainment velocity is relatively small, $v \approx (n_p b_{gb} / b_p) \beta_p (\nabla T)_{matr} << \beta_p (\nabla T)_{matr}$.

For larger pores with $b_p/n_p << b_{gb}$ and with the mobility from Eq. (2a), the condition of coherent motion of the boundary and pores, Eq. (13), takes the form

$$R > \frac{\beta_p (\nabla T)_{matr}}{\pi \gamma n_p b_{gb}}, \tag{14}$$

that is valid (at $T \approx 2273$ K, $(\nabla T)_{matr} \approx 2\cdot 10^5$ K/m) under condition $R > R_{gb}^2 / (n_p \cdot 10^{-3}\,\mathrm{m})$, i.e. for the boundary of the size $R_{gb} \approx 10$ μm the pore radius must exceed $n_p^{-1} \cdot 10^{-7}$ m. In this case

$$v \approx \beta_p (\nabla T)_{matr}, \tag{15}$$

that implies negligible retarding effect for the pores-boundary complex relocation in the temperature gradient practically for the whole range of pore sizes with $R > n_p^{-1} \cdot 10^{-6}$ m .

However, in the case of reduced (by 2 orders of magnitude) value of $D_{gb}$ obtained by interpolation of the correlation measured at lower temperatures, $T \le 1700$°C, the critical pore size becomes much larger, $R > n_p^{-1} \cdot 10^{-5}$ m, i.e. comparable with the grain size. The interval for small pores (defined by $b_p/n_p << b_{gb}$) for this value of $D_{gb}$ correspondingly decreases to $R << n_p^{-1} \cdot 10^{-8}$ m. This implies that intermediate size pores, $n_p^{-1} \cdot 10^{-8}$ m $< R < n_p^{-1} \cdot 10^{-5}$ m, moving in the temperature gradient, detach from grain boundaries and move independently (for this choice of $D_{gb}$). Therefore, it cannot be excluded that only very large pores (comparable with the grain size of ≈ 10 μm and larger) can effectively entrain grain boundaries and thus participate in formation of columnar grains (see Section 2.3).

For effective entrainment of a boundary by pores moving in the temperature gradient, the pore mobility $b_p$ under an external force (controlled by the pore surface diffusion mechanism), has to be much smaller than the boundary mobility. The resulting high velocity of the pores-boundary complex is determined by the high pore mobility $\beta_p$ in the temperature gradient (controlled by the evaporation-condensation mechanism).

In the case of $n_l$ large ($l$) pores and $n_s$ small ($s$) pores attached to the boundary and moving in the temperature gradient by the evaporation mechanism, Eq. (5) is substituted by a pair of equations

$$v_p^{(l)} = \beta_p (\nabla T)_{matr} - b_p^{(l)} F_l, \tag{16a}$$

$$v_p^{(s)} = \beta_p (\nabla T)_{matr} - b_p^{(s)} F_s, \tag{16b}$$

Eq. (7) is reduced to

$$v_{gb} = b_{gb} (n_l F_l + n_s F_s), \tag{17}$$

and Eq. (9) takes the form

$$v = v_{gb} = v_p^{(l)} = v_p^{(s)}. \tag{18}$$

Owing to the relationship $b_p^{(l)} << b_{gb} << b_p^{(s)}$, the velocity of the coherent motion of the complex is determined by large pores behaviour

$$v = \left(1 - \frac{b_p^{(l)} b_p^{(s)}}{b_p^{(l)} b_p^{(s)} + n_l b_{gb} b_p^{(s)} + n_s b_{gb} b_p^{(l)}}\right) \beta_p (\nabla T)_{matr} \approx \left(1 - \frac{b_p^{(l)}}{n_l b_{gb}}\right) \beta_p (\nabla T)_{matr} \approx \beta_p (\nabla T)_{matr}, \tag{19}$$

i.e. Eq. (15) is still valid.

In the case of very large pores, $R >> R_{gr}$, one should take into consideration that a single pore can entrain boundaries of several adjacent grains, the number of which can be evaluated as $N_{gb} \approx (R/R_{gb})^2$. Correspondingly, Eq. (5) takes the form

$$v_p = \beta_p (\nabla T)_{matr} - b_p N_{gb} F \quad , \tag{20}$$

whereas Eqs. (7) and (10) keep their form.

Solution of these equations determines the drag force exerted by the pore on each entrained boundary

$$F = \frac{\beta_p (\nabla T)_{matr}}{N_{gb} b_p + b_{gb}}, \tag{21}$$

and relocation velocity of the grain boundaries attached to the pore

$$v = \frac{b_{gb}}{b_{gb} + N_{gb} b_p} \beta_p (\nabla T)_{matr} . \tag{22}$$

Obtained solutions are valid only under an additional condition $N_{gb} F < F_m$, that takes the form

$$\frac{\beta_p (\nabla T)_{matr}}{b_{gb}/N_{gb} + b_p} < \pi \gamma R . \tag{23}$$

As above indicated, at $R \approx R_{gb}$ (when $N_{gb}$=1) the pore mobility $b_p$ is negligible in comparison with the boundary mobility $b_{gb}$. The pore mobility falls down with increase of its radius proportionally to $R^{-4}$, whereas $N_{gb}$ increases proportionally to $(R/R_{gb})^2$, thus $N_{gb} b_p$ reduces proportionally to $R^{-2}$ and, hence, $N_{gb} b_p$ remains negligibly small in comparison with $b_{gb}$, and Eq. (15) is restored, $v \approx \beta_p (\nabla T)_{matr}$.

This formal result confirms an obvious qualitative conclusion that after attachment to a grain boundary of a very large pore (comparable with or exceeding the grain boundary surface) the boundary practically disappears and thus does not exert any influence on the pore movement.

As a result, Eq. (23) imposes a rather weak restriction to the pore size

$$R < \frac{\gamma}{\beta_p (\nabla T)_{matr}} \frac{D_{gb} \Omega}{wkT}, \tag{24}$$

which under typical conditions $T \approx 2273$ K, $\beta_p (\nabla T)_{matr} \approx 3 \cdot 10^{-9}$ m/s is reduced to $R < 10^{-3}$ м.

As a rule, observed columnar structures in oxide fuels are indeed formed by entrainment of several adjacent boundaries by a large lenticular pore (see, e.g. Fig. 1 from [2]).

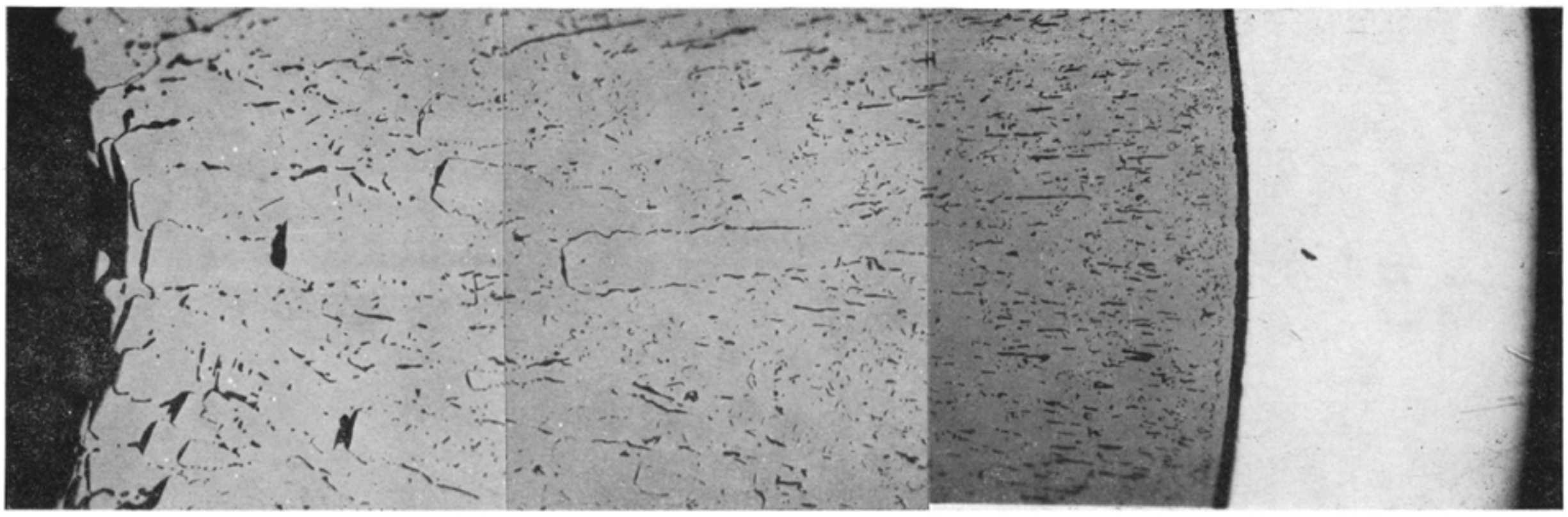

Fig. 1. Cross-section of a $UO_2$ specimen (× 80) irradiated at high temperature in the Battelle Research Reactor (from [2])[1].

## 2.3 Formation of non-equiaxed and columnar grains

The grain boundary velocity in the case of the equiaxed grain growth of non-porous material can be calculated as [13]

$$v_{gb}^{(0)} = b_{gb} \Delta G S , \tag{25}$$

where $\Delta G$ is the driving force for the grain growth (connected with the boundary curvature), $S$ is the boundary surface area. However, in real porous $UO_2$ fuel the grain growth is significantly suppressed [16, 17] and is controlled by migration of pores (in fresh fuel) and/or gas bubbles (in irradiated fuel) attached to grain boundaries [12, 18]. If there are $n_p$ (equisized) intergranular bubbles or pores attached to the boundary, the velocity is reduced owing to the retarding forces exerted on the boundary by the pores [12]

$$v_{gb} = b_{gb} \left( \Delta G S - F n_p \right). \tag{26}$$

In the temperature gradient the attached pores tend to move upward the gradient and produce additional forces exerting on the boundary in the direction of their movement. Under these forces both fore and rear (with respect to the temperature gradient direction) boundaries (moving in the opposite to each other directions) of a growing grain are entrained in the same direction, this can enhance formation of elongated grains.

Indeed, in projection on the direction of the boundary migration (positive for the fore boundary and negative for the rear one) the velocity of a pore attached to the corresponding boundary of the growing grain is determined by superposition of the force exerted by the boundary and the driving force induced by the temperature gradient

$$v_p = b_p F \pm \beta_p (\nabla T)_{matr} . \tag{27}$$

Solution of Eqs. (26), (27) and (10) determines velocities of the fore and rear boundaries (migrating in the opposite directions)

$$v_{gb}^{(+,-)} = \frac{b_{gb} \left( b_p / n_p \right)}{b_{gb} + \left( b_p / n_p \right)} \Delta G S \pm \frac{b_{gb}}{b_{gb} + \left( b_p / n_p \right)} \beta_p (\nabla T)_{matr} \approx v_{gb}^{(0)} \pm \beta_p (\nabla T)_{matr} . \tag{28}$$

---

[1] This paper article published by W. Chubb, V. W. Storhok and D. L. Keller in J. Nucl. Mater. 44 (1972) 136-152, Copyright Elsevier

Therefore, the mean velocity of the grain elongation in the direction of the temperature gradient can be roughly evaluated as

$$\frac{dL}{dt} = v_{gb}^{(+)} + v_{gb}^{(-)} \approx \left(\frac{d\left(2\bar{R}_{gr}\right)}{dt}\right)_0 + \left[\left(\nabla\beta_p\right)\left(\nabla T\right) + \beta_p\left(\nabla^2 T\right)\right]L$$
$$= \left(\frac{d\left(2\bar{R}_{gr}\right)}{dt}\right)_0 + \left[\frac{d\beta_p}{dT}\left(\nabla T\right)^2 + \beta_p\left(\nabla^2 T\right)\right]L, \quad (29)$$

where the first term in the r.h.s. of Eq. (29) determines the equiaxed growth velocity, calculated as [19, 20]

$$\left(\frac{d\bar{R}_{gr}}{dt}\right)_0 \approx 0.1\frac{2\gamma_{gb}}{\bar{R}_{gr}}\frac{b_{gb}}{1 + b_{gb}\sum n_p(R) b_p^{-1}(R)}, \quad (30)$$

whereas the second term determines elongation of the growing grain. Since the spatial distribution of temperature in the fuel pellet is close to parabolic, the term $\left(\nabla^2 T\right)$ weakly varies in the radial direction and is negative (as well as the term $\left(\nabla T\right)$), whereas the term $\left(d\beta_p/dT\right)$ is positive and thus determines a positive input in the elongation velocity. It can be estimated that at $T \approx 2273$ K and $\nabla T \approx 2\cdot 10^5$ K/m the term $\alpha \equiv \frac{d\beta_p}{dT}\left(\nabla T\right)^2$ is of the order of $\approx 10^{-5}$ s$^{-1}$ and $\alpha >> \left|\beta_p\left(\nabla^2 T\right)\right|$ (outside the inner cell, where $\nabla T$ tends to zero).

In the initial stage of irradiation when the first term in Eq. (29) dominates, $\left(\frac{d\left(2R_{gr}\right)}{dt}\right)_0 >> \alpha L \approx 2R_{gr}\alpha$, grains grow as almost equiaxed (with some possible elongation owing to different velocities of fore and rear grain boundaries moving at different temperatures). However, the grain growth velocity rapidly decreases and after some time (≤ $10^4$ s, depending on temperature) the term $\alpha L$ that determines the grain elongation, commences to dominate in Eq. (29). This qualitatively explains the observed fuel microstructure with elongated grains induced by the temperature gradient and sintering pores in the hot zone of pellets, which was formed in the first hours of irradiation, see Fig. 2 from [21].

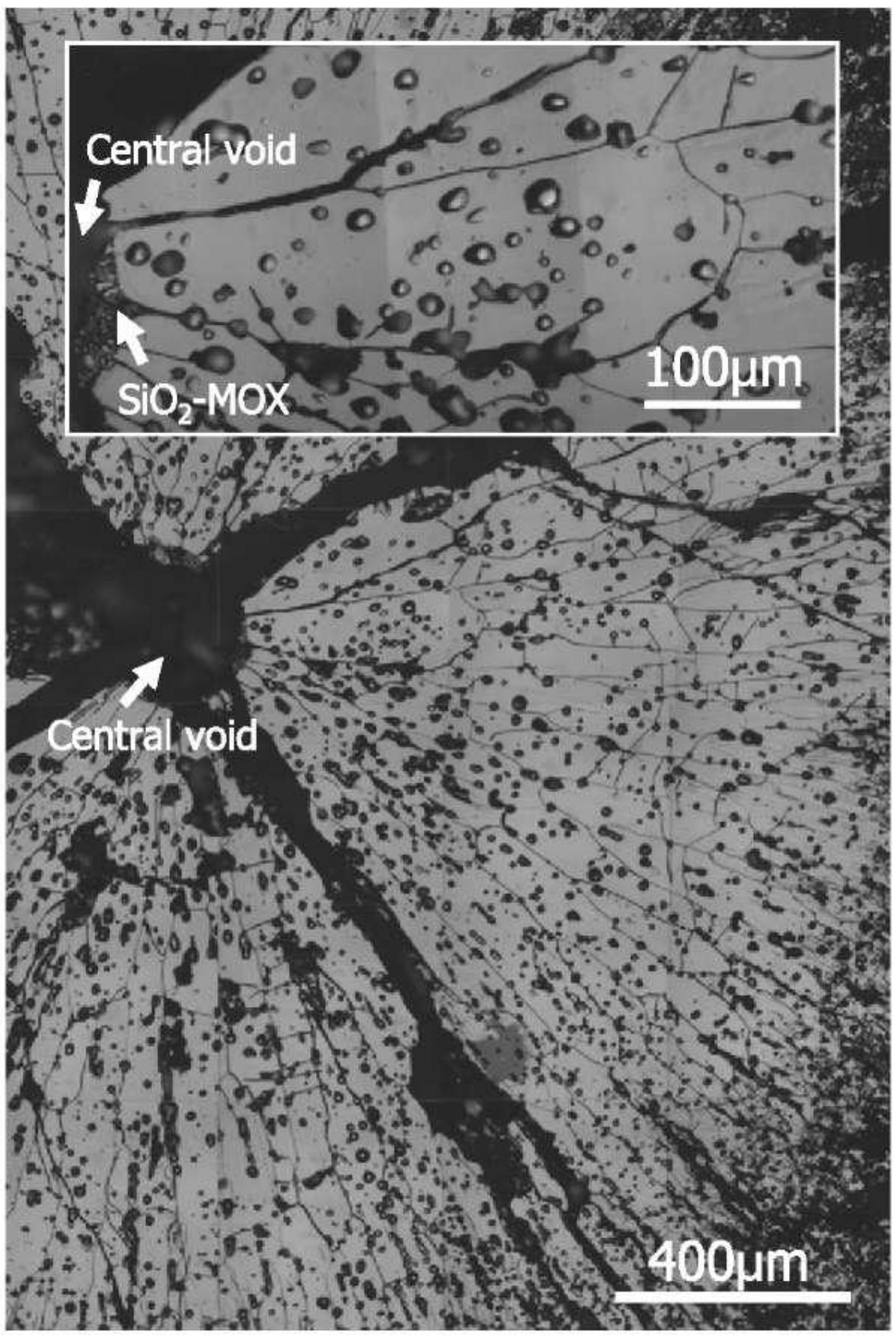

Fig. 2. Fuel microstructure of the chemically etched oxide fuel in the experimental fast reactor Joyo after continuous power increase within the first ≈ 5 hours of reactor operation and the subsequent 10 min. at the maximum power level with the linear heating rate of ≈ 430 W/cm (from [21])[2].

In the subsequent period of irradiation (≈ $10^5$ s) sintering pores intensively shrink in the central hot zone of fuel pellets (2000-2500 K), in accordance with results of pore shrinkage rate measurements in this temperature range [22] and with direct observations [23] of a notable fuel densification in the pellet central part during the first ≈ 300 hours of fast reactor operation (also observed in [2]). This conclusion can be also confirmed by calculations (see below Section 3). During this period the majority of pores shrink (and transform into gas filled bubbles), with exception of very large pores.

However, intergranular gas-filled bubbles of (sub)micron size that appear under irradiation conditions, have much smaller mobility in the temperature gradient in comparison with comparable size pores. In fact, since the gas pressure in small bubbles $P_b \approx 2\gamma/R_b$ is much higher than that in pores (≈ 1 bar), the vapour diffusivity $D_{fg}$ in bubbles is much lower than that in pores, and the migration velocity of bubbles in the temperature gradient is correspondingly relatively low (in comparison with pores of comparable size). Besides, a relatively small geometry factor $\chi$ in Eq. (2) of (close to spherical) survived sintering pores ($\chi \approx 0.5$) in comparison with that of lenticular pores ($\chi \geq 3$), also significantly reduces their velocity in comparison with that of lenticular pores.

---

[2] This paper article published by K. Maeda, S. Sasaki, M. Kato, Y. Kihara in J. Nucl. Mater. 385 (2009) 413–418, Copyright Elsevier

Therefore, the columnar structure formation is controlled by migration of large lenticular pores (see Fig. 1). Indeed, since concentration of large lenticular pores is very small in comparison with concentration of grains in fuel pellets, the probability of simultaneous capture of two large lenticular pores on the opposite boundaries of a grain is also small. For this reason, a lenticular pore captured by a boundary entrains this boundary with the velocity significantly higher than the velocity of the opposite boundary controlled by migration of (much slower) small intergranular bubbles and/or pores. This results in formation of columnar grains.

Numerical calculations of velocity of grain boundaries entrained by large lenticular pores, using Eqs. (2) and (15) with typical for fast reactors temperature distribution, are presented in Fig. 3. Depending on the initial position of a pore, entrained boundaries will migrate during different time intervals, Fig. 4, and to different distances, Fig. 5, before their velocity will strongly reduce in the innermost part of a pellet.

From these figures one can evaluate that the length of the formed grains attains ≈ 1-2 mm (compare with Fig. 1), whereas the size of the columnar grains zone attains a half of the pellet radius during the initial period of ≈ 100 hours, in a fair agreement with numerous observations (see, e.g. [23]).

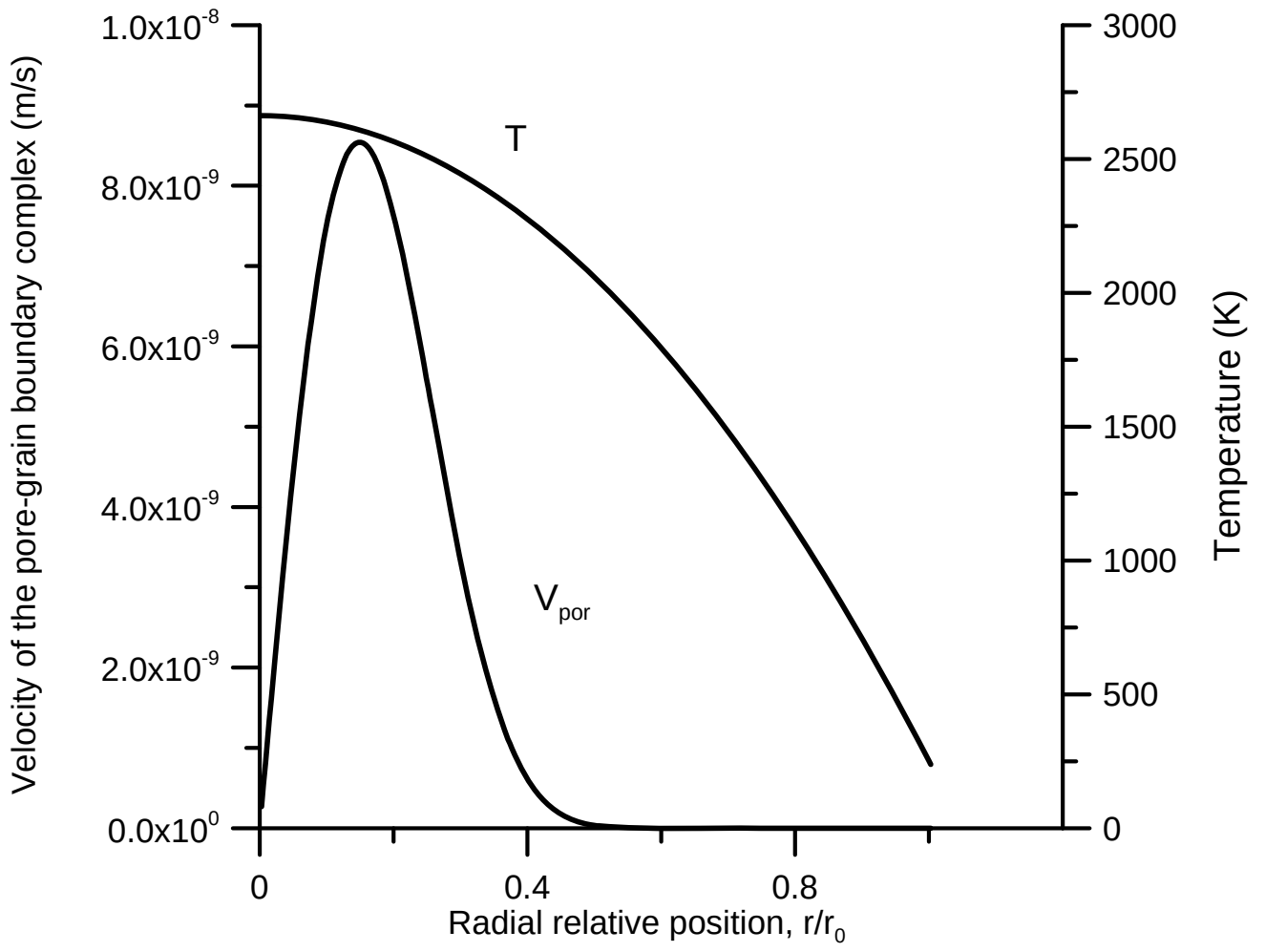

Fig. 3. Calculated radial distribution of the pore-boundary complex velocities for a typical temperature profile in the fast reactor fuel pellet (of radius $r_0$).

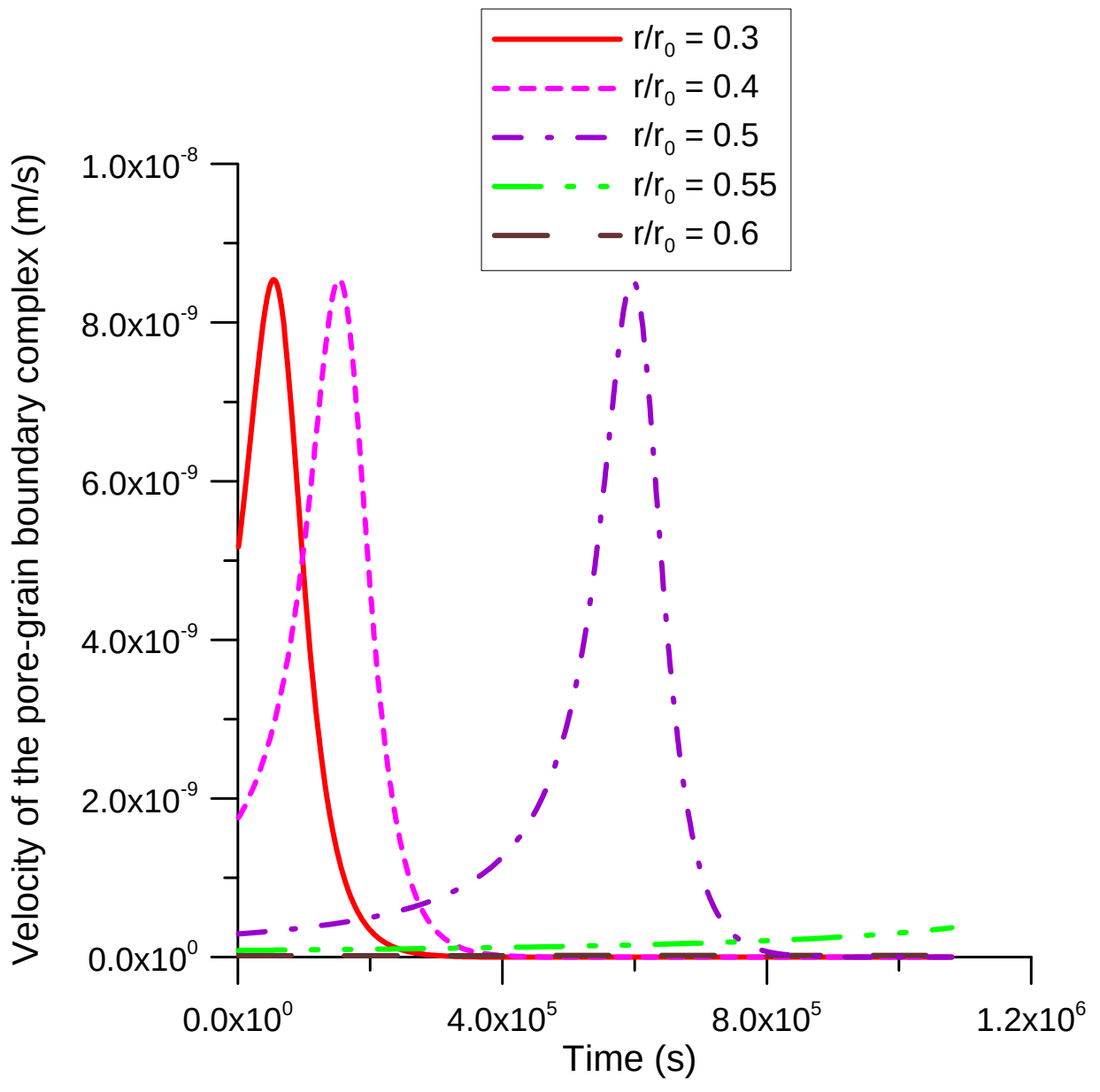

Fig. 4. Calculated variation of the pore-boundary complex velocity with time for different initial radial relative positions in the fuel pellet (of radius $r_0 = 3$ mm).

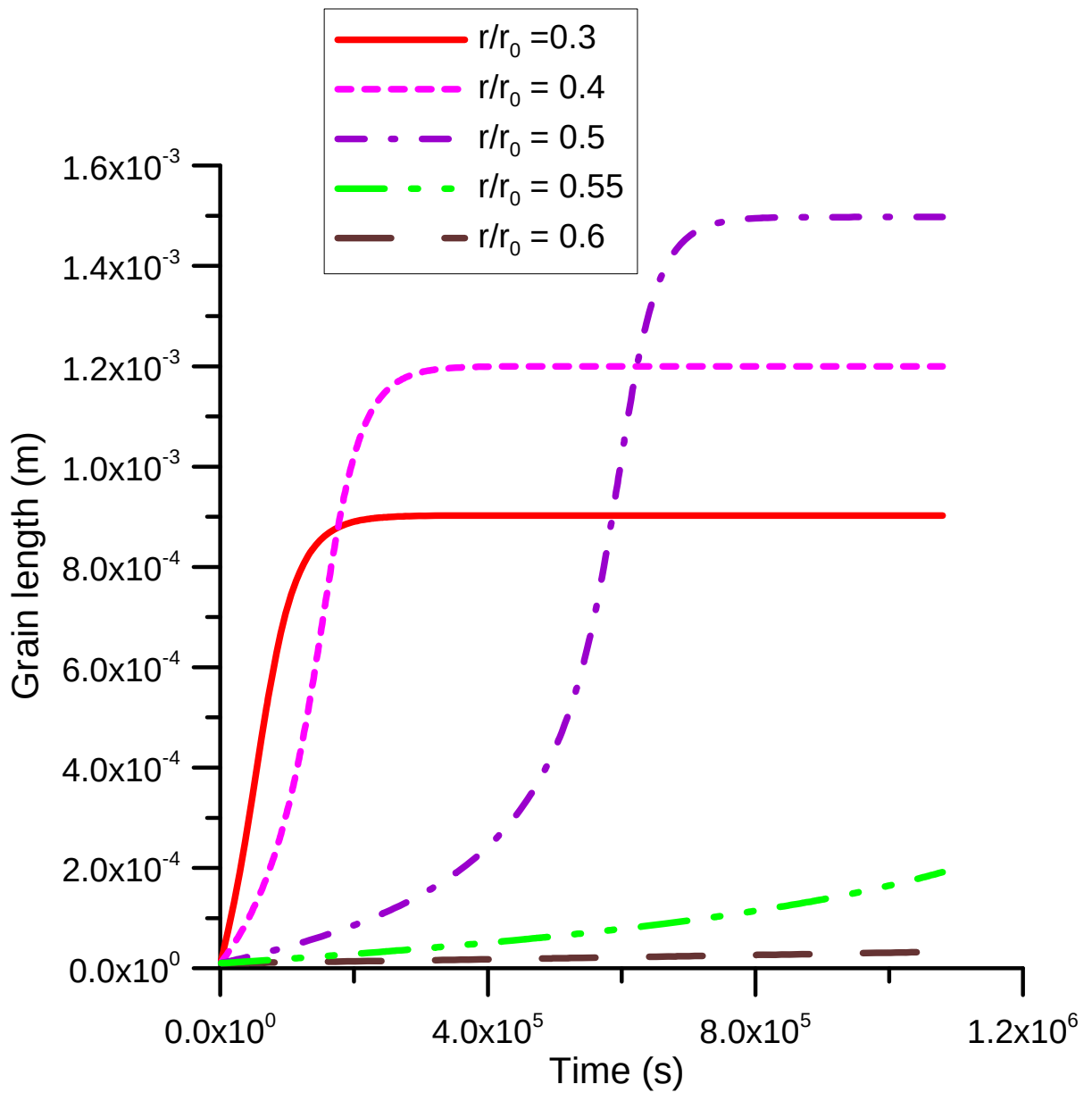

Fig. 5. Calculated growth of columnar grains, located in various radial relative positions in the fuel pellet (of radius $r_0 = 3$ mm).

# 3 Model for central void growth (initial stage of irradiation)

Lenticular pores have also been assumed to collect at the centre line of the fuel, giving rise to the formation and/or growth of a central cavity [5]. In order to explain quantitatively a relatively high growth rate observed in reactor tests, it was additionally assumed that all as-fabricated pores participate in this process [24, 25].

However, in a cylindrical fuel rod with an axisymmetric power distribution, the thermal gradient is practically zero at the centre of the pin or inner surface if a central void exists (owing to zero heat flux condition). Therefore, the pore velocity (proportional to the temperature gradient) is smoothly reducing in the central zone and turns to zero in the centre (or near the void surface), see Fig. 3.

In the case of large pores, which cannot be considered as point-wise objects, one should take into consideration different values of $T$ and $\nabla T$ at the fore and rear surfaces of a pore and calculate their velocities (using similar to Eq. (2) expressions) separately (cf. [4]). This will apparently result in additional deformation of the pore, whereas velocity of the pore advancement to the centre will be determined by position of the pore fore surface, which tends to zero near the central void surface. This significantly retards pores sinking into the void and results in accumulation of porosity in the innermost cell during a relatively long initial period ($\geq 10^5$ s) of reactor irradiation.

It should be also noted that thermoelastic stresses in fuel induced by the temperature gradient also tend to zero at the central void surface (along with the temperature gradient) and thus cannot accelerate relocation of pores from the innermost cell to the central void, since the relocation velocity of a pore under external stresses is proportional to the stress gradient, $v_p = (5/3)(D_{vol}/kT)\nabla\sigma$ or $v_p = (10/3)(D_s\Omega\delta/kTR)\nabla\sigma$ (for two different migration mechanisms, cf. Eq. (6) ) [9].

To overcome this situation, the thermal gradient at the inner fuel cell boundary is often considered to be the average value across the central cell (as explicitly postulated in [24] and, probably, implicitly assumed in [25]). In order to avoid this non-physical assumption, a new, complementary mechanism of the central void growth is proposed in the current Section.

In accordance with numerous observations, a notable densification takes place in the central zone of fuel pellets during some initial period ($10^5$ – $10^6$ s) of fast reactor operation. The pores shrinkage model [26] of the MFPR code [27, 28], initially developed for moderate temperature conditions of LWRs (where shrinkage is mainly determined by irradiation-induced rather than by thermal processes) and then modified for consideration of thermally activated densification [29], was thoroughly verified against microscopic measurements [22] of porosity evolution and applied (along with the fuel swelling model of the code) to consideration of high-temperature densification in the central zone of pellets in fast reactors, Fig. 6 (from [29]).

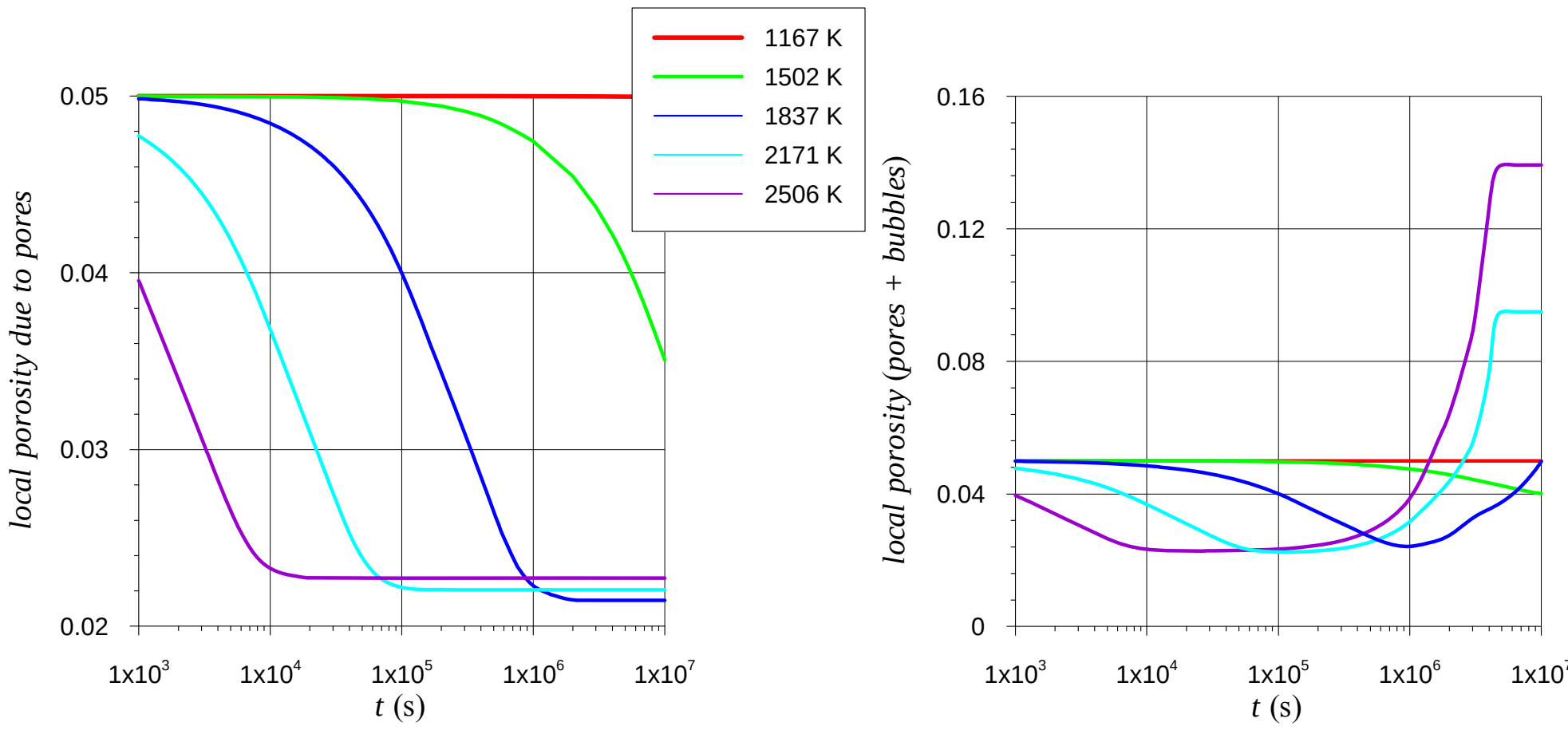

Fig. 6. MFPR simulation of local porosity evolution with time in $UO_2$ fuel at different temperatures owing to sintering pores shrinkage and intergranular bubbles growth under typical fast reactor irradiation conditions (from [29]).

As seen from Fig. 6, the model predicts a notable fuel densification in the inner zone and negligible effect in the (cooler) outer zone of pellets within the first $10^5$ – $10^6$ s, in a qualitative agreement with direct observations of fuel behaviour after 300 hours of irradiation in DUELL tests performed in the high-flux reactor HFR (Petten) [23]. Model predictions of residual porosity of ≈ 2% (in the left panel of Fig. 6) are also in a good agreement with observations in depleted zone of $UO_2$ pellets (i.e. in the lack of fission gas bubbles) at temperatures 1400-1600°C after ≈ $10^7$ s of irradiation in the Battelle Research Reactor tests [2].

Owing to inhomogeneous fuel densification, with a notably increased density in the inner zone and (close to) as-fabricated density in the outer zone of the pellet, tensile (positive) stresses are induced in the inner zone, smoothly reducing in radial direction to zero and converting to compressive (negative) stresses in the outer zone. Under these stresses high-temperature creep is activated in the pellet hot inner zone, sustaining radial mass transport from the pellet centre in radial direction that results in the central void growth.

Lacking in the code a mechanistic fuel creep model, a conservative approach can be applied assuming that the local pore shrinkage in a grain (by vacancy diffusion from the pore to the grain boundaries) is the rate controlling process rather than large-scale mass transport by creep. This assumption is analogous to the traditional approach to consideration of the fuel pellet swelling, where the local pore growth in a grain (by vacancy diffusion from the grain boundaries to the pore) is assumed to be the rate controlling process, whereas the large-scale radial mass transport (by creep) from these grains (located all over the pellet) to the fuel pellet periphery (that provides outward expansion of the pellet) is a rapid process.

Under this assumption the central cavity growth can be determined from the mass (or volume) conservation law

$$\left(1-\overline{P}(t)\right)\left(R_O^2(t)-R_h^2(t)\right)=\left(1-P_0\right)\left(R_O^2(0)-R_I^2(0)\right), \tag{31}$$

where $R_h$ is the central void radius, $R_O$ and $R_I$ are the pellet outer and inner radii, respectively; $P_0$ is the as-fabricated porosity, $\overline{P}$ is the mean porosity varied owing to pores shrinkage (fuel densification).

A formally similar to Eq. (31) volume balance equation is also used in the above discussed models for the central void growth [23 - 25], however, with a different physical sense. In fact, variation of $\overline{P}(t)$ in these models is assumed taking place owing to pores sinking into the central void as a result of their migration in the temperature gradient, rather than owing to pores shrinkage, as assumed in the current approach. Besides the problem of pores accumulation in the innermost cell (retarding their sinking into the void) in the traditional approach, capturing of relocating pores by grain boundaries and possible retarding of their motion (as discussed in Section 2) cannot be generally ignored (for small and intermediate-size pores); this may additionally complicates description of the central void growth rate by this mechanism.

It should be noted that, if pores cannot reach the central void surface (and thus cannot directly participate in its growth), their accumulation in the innermost cell, which has the hottest temperature, accelerates thermal shrinkage of these pores and thus additionally increases densification rate (and thus the creep rate) of the pellet. For this reason, one can conclude that migration of pores enhances (indirectly) the central void growth rate. However, taking into consideration that notable relocation of pores occurs in the inner half-radius zone of the pellet with T ≥ 2000 K (see Fig. 5), where densification process is practically completed within the initial time period of ≈ $10^5$ s independently of their position in the inner zone (see Fig. 6), this effect can be neglected in the first, conservative approximation. Nevertheless, this process of pores migration to the pellet hot zone and their accelerated shrinkage in this zone can be self-consistently considered in an advanced model (as foreseen in the future development of the MFPR code, along with implementation of the fuel pellet creep model).

Moreover, trapping of pores accumulated in the innermost cell by the growing central void cannot be ruled out as well, which will further accelerate resulting growth rate of the central void. This may substantiate the traditional mechanism (supplementing it with the final step of pores sinking into the central void), which can be properly taken into consideration in the advanced model, including both models (current and traditional) as complementary mechanisms.

Nevertheless, as above mentioned, observations in DUELL tests [23] reasonably confirm a flat total porosity profile of ≈ 2% attained in the inner half-radius zone after ≈ 300 hours of fast reactor irradiation, this fairly justifies applicability of the above described conservative approach predicting similar to these observations behaviour.

In this approach, calculation of the central void growth is performed with the current version of MFPR conservatively using Eq. (31) combined with the fuel densification model, Fig. 7. For fuel pellets of radius 3 mm with the central void of radius 0.9 mm and material porosity of 6.3% in BN-600 fast reactor [30] the model predicts increase of the cavity radius by ≈ 200 μm, whereas for the solid-core fuel pellet of radius 3.3 mm with the initial porosity of 3.5% in HFR reactor under conditions of DUELL tests [23] (linear power ≈ 450-500 W/cm, evaluated fuel centreline temperature ≈ 2250-2300°C) the model predicts formation of the central hole with the radius of ≈ 500 μm within the first ≈ 300 hours (when essential densification takes place), in a sound agreement with observations.

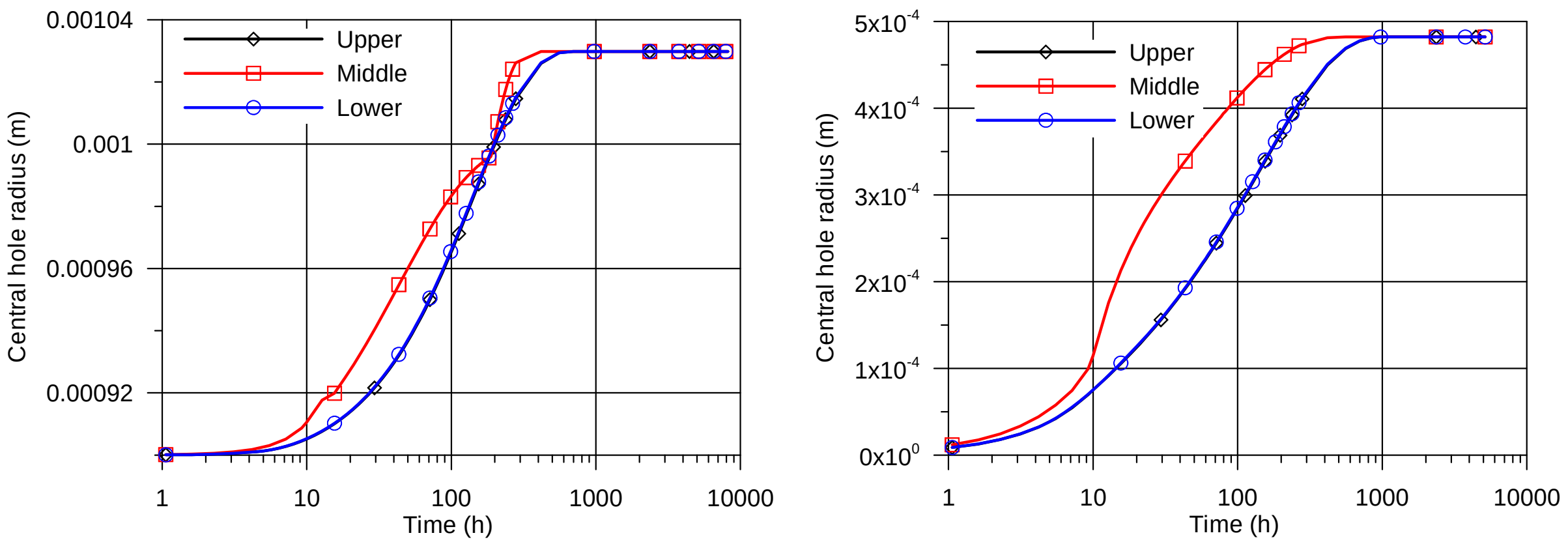

Fig. 7. MFPR simulation of the central cavity growth at three axial levels of fuel rods, owing to radial mass transport in the fuel pellet induced by high-temperature creep in the fuel pellet of BN-600 fast reactor with the initial cavity radius of 0.9 mm, pellet radius of 3 mm and porosity of 6.3% (*left*) and in the solid-core fuel pellet of radius 3.3 mm of HFR reactor (Petten) with the initial porosity of 3.5% (*right*).

In the subsequent period of irradiation (≥ $10^6$ s) when the total porosity commences to increase owing to gas bubbles growth (see Fig. 6), the central void shrinkage will occur in competition with pellet swelling in the outward direction, so, it will be not strong and can be properly described by the creep model with self-consistent consideration of induced stresses and mass transfer in both (inward and outward) radial directions (as foreseen in the future development of the MFPR code).

# 4 Conclusions

On the base of analysis of experimental observations and critical assessment of existing models for oxide fuel structure evolution under operation conditions of fast reactors, new models for fuel restructuring and coring are proposed.

The restructuring model describes coherent motion of grain boundaries with attached pores in the temperature gradient. The mobility of pores in a high temperature gradient is extremely high, owing to the effective evaporation-condensation mechanism under the temperature gradient across the pore. On the other hand, their mobility under external forces (exerted by grain boundaries) can be very low in comparison with the inherent mobility of the boundaries to which pores are attached. As a result of self-consistent consideration of these effects (low mobility of pores under external forces and their high mobility in the temperature gradient), it is shown that the retarding effect of a boundary on a pore is negligible in the case of large pores. Therefore, the velocity of a pore-grain boundary complex practically coincides with the velocity of a free pore migrating in the temperature gradient. This justifies validity (in the case of large pores) of the simplified traditional approach that ignores consideration of mutual interactions between grains and pores. The model explains elongation of thermally growing equiaxed grains (by bubbles and sintering pores) and formation of columnar grains (by large lenticular pores), owing to entrainment of grain boundaries by the attached pores, and allows calculating a rapid formation of extended zone ($\geq 0.5$ of pellet radius) with very large (up to 1-2 mm length) columnar grains during the first $\approx 100$ hours of fast reactor irradiation, in a reasonable agreement with experimental observations.

The coring model describes formation and growth of the central void in the pellet. It is shown that the traditional models, which assume collecting of pores at the centre line of the fuel owing to their migration in the temperature gradient, are inconsistent with the zero heat flux condition at the centre line (or at the central void surface) that prevents the pores from rapid sinking into the central void. The new model explains the coring by the radial mass transport from the pellet centre to the periphery by the creep mechanism and additionally supplement the traditional mechanism with the final step in relocation of pores (accumulated in the innermost cell) into the central void. The creep mechanism operates under tensile stresses induced in the pellet central zone owing to inhomogeneous fuel densification, connected with accelerated thermal shrinkage of sintering pores in the hot inner zone during the initial period ($\leq 300$ hours) of fast reactor operation.

A simplified (conservative) approach (similar to the traditional consideration of fuel swelling) is applied that is valid under assumption that the local pore shrinkage in a grain (by vacancy diffusion from the pore to the grain boundaries) is the rate controlling (slow) process, whereas the radial mass transport (by creep mechanism) from these densifying grains (located in the hot inner zone of the pellet) to the outer zone (with as-fabricated density) is a rapid process. Being implemented in the MFPR code along with the pore shrinkage model, the model allows conservative predictions for the central void growth rate (during the initial stage of irradiation, when fuel densification takes place) in a reasonable agreement with observations.

It is foreseen to improve the central void growth model in future development of the MFPR code by implementation of the fuel pellet creep model and additional consideration of accelerated densification of pores owing to their relocation (in the temperature gradient) to the hot inner zone of the pellet and subsequent sinking into the central void owing to counter motion of the growing cavity surface.

## Acknowledgements

The author thanks Dr. V.I. Tarasov (IBRAE) for kind delivery of his calculation results (Fig. 6), for careful reading of the manuscript and for valuable comments, Dr. V.E. Shestak (IBRAE) for his support with MFPR calculations (Fig. 7), Mr. I.B. Azarov (IBRAE) for his support with numerical calculations (Figs. 3-5) and Dr. A.V. Boldyrev (IBRAE) for numerous valuable discussions. This work was supported by the Russian Federally Targeted Program on New

Technology Platform and by the Russian Foundation for Basic Research, which are greatly acknowledged by the author.